\documentclass{article}
\usepackage{amssymb}

\usepackage{fleqn}
\usepackage{psfig}
\usepackage{epsfig}
\usepackage{graphicx}
\usepackage{amsmath}

\input{tcilatex}
\def\beq{\begin{equation}}
\def\eeq{\end{equation}}
\def\bea{\begin{eqnarray}}
\def\eea{\end{eqnarray}}
\def\bq{\begin{quote}}
\def\eq{\end{quote}}

\def\gappeq{\mathrel{\rlap {\raise.5ex\hbox{$>$}}
{\lower.5ex\hbox{$\sim$}}}}
\def\lappeq{\mathrel{\rlap{\raise.5ex\hbox{$<$}}
{\lower.5ex\hbox{$\sim$}}}}
\def\Toprel#1\over#2{\mathrel{\mathop{#2}\limits^{#1}}}

\begin{document}

\pagestyle{empty} 
\begin{flushright}
{CERN-TH/2001-099}\\
hep-ph/0104020\\
\end{flushright}
\vspace*{15mm}

\begin{center}
\textbf{\\[2pt]
RESUMMED $B\rightarrow X_{u}\,l\,\nu \,$ DECAY DISTRIBUTIONS TO
NEXT-TO-LEADING ORDER } \\[9pt]
\vspace{2cm} $~~$\\[0pt]
\textbf{U. Aglietti}$^{\ast )}$ \\[0pt]
\vspace{0.1cm} $~~$\\[0pt]
Theoretical Physics Division, CERN\\[1pt]
CH - 1211 Geneva 23 \\[3pt]
$~~$ \\[0pt]
\vspace*{1.5cm} \textbf{Abstract} \\[0pt]
\end{center}

We perform factorization of the most general distribution in semileptonic $%
B\rightarrow X_{u}$ decays and we resum the threshold logarithms to
next-to-leading order. From this (triple-differential) distribution, any
other distribution is obtained by integration. As an application of our
method, we derive simple analytical expressions for a few distributions,
resummed to leading approximation. It is shown that the shape function can
be directly determined by measuring the distribution in $%
m_{X}^{2}/E_{X}^{2}, $ not in $m_{X}^{2}/m_{B}^{2}.$ We compute the resummed
hadron energy spectrum, which has a ``Sudakov shoulder'', and we show how
the distribution in the singular region is related to the shape function. We
also present an improved formula for the photon spectrum in \textbf{$%
B\rightarrow X_{s}\,\gamma ,$} which includes soft-gluon resummation and
non-leading operators in the effective hamiltonian. We explicitly show that
the same non-perturbative function --- namely the shape function ---
controls the non-perturbative effects in all the distributions in the
semileptonic and in the rare decay. \vspace*{3cm} \noindent 
%\rule[.1in]{16.5cm}{.002in}

\noindent $^{\ast )}$ On leave of absence from Dipartimento di Fisica,
Universit\`{a} di Roma I, Piazzale Aldo Moro 2, 00185 Roma, Italy. E-mail
address: ugo.aglietti@cern.ch. \vspace*{0.5cm}

\begin{flushleft} CERN-TH/2001-099 \\[0pt]
April 2001
\end{flushleft}
\vfill\eject
%\pagestyle{empty}
%\clearpage\mbox{}\clearpage

\setcounter{page}{1} \pagestyle{plain}

\bigskip

\section{Introduction}

In this note we discuss factorization and threshold resummation to
next-to-leading order (NLO) in semi-inclusive $B$ decays: 
\begin{equation}
B\rightarrow X_{q}+\mathrm{(non\,\,\,QCD\,\,\,partons),}  \label{one}
\end{equation}
where $X_{q}$ is any hadronic final state containing the light quark $q=u,s$
coming from $b$ fragmentation. The non-QCD partons are a lepton--neutrino
pair or a photon.

In sec.\thinspace 2 we consider the most general distribution in the decay 
\begin{equation}
B\rightarrow X_{u}+l+\nu .  \label{sldec}
\end{equation}
A general expression for the triple differential distribution is presented,
from which any other distribution is obtained by integration. The process (%
\ref{sldec}) is characterized by three energy scales: 
\begin{equation}
m_{B},\quad Q,\quad m_{X},
\end{equation}
where\footnote{%
The factor $2$ multiplying $E_{X}$ is inserted for convenience.} 
\begin{equation}
Q\equiv 2E_{X}.
\end{equation}
Factorization is achieved with two different steps, according to the method
developed in \cite{congiulia2}:

\begin{enumerate}
\item  We factorize heavy mass effects by taking the limit 
\begin{equation}
m_{B}\rightarrow \infty ,\qquad Q,\,\,\,\,m_{X}\rightarrow \mathrm{const.}
\label{limite1}
\end{equation}
The distributions contain logarithms of the heavy mass, $\log m_{B}/Q,$
related to the non-conservation of the transition currents in the static
limit for the beauty quark \cite{dimenticato}. This step is discussed in
sec.\thinspace 2.1.

\item  We factorize the large infrared logarithms that appear in the
semi-inclusive region, 
\begin{equation}
\alpha _{S}^{\,n}\,\mathcal{D}_{k}\left( z\right) \qquad \left( 0\leq k\leq
2n-1\right) ,
\end{equation}
by taking the limit 
\begin{equation}
z\rightarrow 1  \label{limite2}
\end{equation}
where 
\begin{equation}
z\equiv 1-\frac{m_{X}^{2}}{Q^{2}}.  \label{zetavar}
\end{equation}
We have defined: 
\begin{equation}
\mathcal{D}_{k}\left( z\right) \equiv \left[ \frac{\log ^{k}\left(
1-z\right) }{1-z}\right] _{+}.
\end{equation}
Plus-distributions are defined as usual as $P\left( z\right) _{+}\equiv
P\left( z\right) \,-\,\delta \left( 1-z\right) \int_{0}^{1}dx\,P\left(
x\right) .$ This step is discussed in sec.\thinspace 2.2.
\end{enumerate}

The basic idea of our approach is to use the kinematical variables 
\begin{equation}
w\equiv \frac{Q}{m_{B}}\quad \left( 0\leq w\leq 2\right) \qquad \mathrm{and}%
\qquad z,
\end{equation}
the latter replacing the generally used one 
\begin{equation}
u\equiv 1-\frac{m_{X}^{2}}{m_{B}^{2}}.
\end{equation}
In other words, we consider (two times) the hadronic energy $Q$ as the hard
scale of the process, rather than the $b$ quark mass $m_{b}$ (or,
equivalently, the $B$-meson mass). The reason for this choice is that the
infrared structure of the decay is not modified by the limit (\ref{limite1}%
): this implies that $m_{B}$ cannot be the relevant hard scale; the heavy
mass acts only as an energy resevoir for the light partons in the final
state and has not any fundamental dynamical meaning. Using the variable $z$
instead of $u$ largely simplifies the logarithmic structure. In the latter
variable, the decay distributions contain, at one loop, terms of the form 
\cite{ndf}: 
\begin{equation}
\mathcal{D}_{1}\left( u\right) ,\quad \quad \log w\,\mathcal{D}_{0}\left(
u\right) ,\quad \quad \mathcal{D}_{0}\left( u\right) ,\quad \quad \log
^{2}w\,\delta \left( 1-u\right) ,\quad \quad \log w\,\delta \left(
1-u\right) ,
\end{equation}
while in the variable $z$\ one has only terms of the form: 
\begin{equation}
\mathcal{D}_{1}\left( z\right) ,\quad \quad \mathcal{D}_{0}\left( z\right)
,\quad \quad \log w\,\delta \left( 1-z\right) .
\end{equation}
In the latter case, the two different logarithmic structures basically
decouple.

We explicitly show that a unique function $f\left( z\right) ,$ depending
only on $z,$ factorizes the long-distance effects in any distribution in the
decay (\ref{sldec}). The perturbative expansion of $f\left( z\right) $
reads: 
\begin{equation}
f(z)\equiv \delta \left( 1-z\right) -\alpha _{S}\,A_{1}\,\mathcal{D}%
_{1}\left( z\right) +\alpha _{S}\,B_{1}\,\mathcal{D}_{0}\left( z\right)
+O(\alpha _{S}^{2}),  \label{perturb}
\end{equation}
where the constants multiplying the distributions are given by: 
\begin{equation}
A_{1}=\frac{C_{F}}{\pi },\qquad B_{1}=-\frac{7}{4}\frac{C_{F}}{\pi }
\end{equation}
and $C_{F}=(N_{c}^{2}-1)/(2N_{c})=4/3.$ This function is related, by a
short-distance coefficient function $C,$ to the shape function in the
effective theory \cite{nostri,generale}, defined as 
\begin{equation}
\varphi \left( k_{+}\right) \equiv \langle B\left( v\right) |h_{v}^{\dagger
}\,\delta \left( k_{+}-iD_{+}\right) \,h_{v}|B\left( v\right) \rangle ,
\label{shape}
\end{equation}
where 
\begin{equation}
k_{+}\equiv -\frac{m_{X}^{2}}{Q}.
\end{equation}
$C$\ is defined by the relation \cite{penultimo,ultimo}: 
\begin{equation}
\varphi \left( k_{+};Q\right) ^{QCD}=\int C\left( k_{+}-k_{+}^{\prime
};Q,\mu \right) \,\varphi \left( k_{+}^{\prime };\mu \right)
\,dk_{+}^{\prime },  \label{coefun}
\end{equation}
where we have defined 
\begin{equation}
\varphi \left( k_{+};Q\right) ^{QCD}=\frac{1}{Q}\,f\left( z\right)
\end{equation}
and $\mu <Q$ is the ultraviolet cut-off or renormalization point of the
effective theory. The shape function factorizes the long-distance effects
--- both perturbative and non-perturbative --- in the process up to the
scale $\mu .$ The relation between the QCD and the effective theory variable
is 
\begin{equation}
z=1+\frac{k_{+}}{Q}.
\end{equation}
The relevance of the variable $z$ to describe the long-distance effects ---
perturbative and non-perturbative --- in (\ref{one}) is also supported by
the following argument. From dimensional analysis, the shape function is of
the form 
\begin{equation}
\varphi \left( k_{+};\mu \right) =\frac{1}{\mu }\,\psi \left( \frac{k_{+}}{%
\mu }\right) .
\end{equation}
The coefficient function in eq.\thinspace (\ref{coefun}) does not contains
large logarithms of the ratio $Q/\mu $ if we choose 
\begin{equation}
\mu \sim O\left( Q\right) ,
\end{equation}
implying that 
\begin{equation}
\frac{k_{+}}{\mu }\sim -\frac{m_{X}^{2}}{Q^{2}}=-1+z.
\end{equation}
The last equation implies that long-distance effects are described by a
function of the variable $z,$ Q.E.D.. Let us stress that $f(z)$ ``includes''
the shape function but does not coincide with the latter, because it also
contains some short-distance effects (soft gluons with energies bewteen $\mu 
$ and $Q$ and hard collinear contributions) \cite{penultimo,ultimo}.

The shape function was originally derived with an Operator Product Expansion
(OPE) and was claimed to be the universal non-perturbative component in (\ref
{one}) since the early papers on the subject \cite{nostri,generale}: we
present here an explicit proof of this. On the perturbative side, our
analysis can be considered as the resummed analogue of the distributions
computed to $O\left( \alpha _{S}\right) $ in \cite{ndf}; some of the decay
distributions presented in \cite{ndf} were computed before in \cite
{kuhn,limetto,civuole} \footnote{%
Analogous computations were performed a long time ago in the context of
one-loop QED corrections to $\mu $ decay \cite{sirlin}.}.

In the second part of sec.\thinspace 2 we present simple analytical
expressions for a few distributions in (\ref{sldec}), resummed to leading
logarithmic accuracy. An interesting distribution from the theoretical side
is the hadron energy spectrum. This distribution has a singularity in
fixed-order perturbation theory inside the physical region, in the point 
\cite{ndf} 
\begin{equation}
E_{X}=\frac{m_{B}}{2}.  \label{poinsing}
\end{equation}
This phenomenon originates from the mismatch of the allowed phase space in
lowest order in $\alpha _{S}$:\footnote{%
Kinematics involve the decay of a massive particle into three massless
particles and it is analogous to the well-known case $e^{+}e^{-}\rightarrow
\gamma ^{\ast },Z^{\ast }\rightarrow q\overline{q}g.$} $E_{X}\leq m_{B}/2,$
and in higher orders: $E_{X}\leq m_{B}.$ The all-order resummation
eliminates the singularity and produces a characteristic behaviour known as
the ``Sudakov shoulder'' \cite{sudshoul}. Actually, a singularity very close
to point (\ref{poinsing}) remains even after soft-gluon resummation. The
latter has a different origin: it is related to the Landau pole in the
running coupling and is factorized with the introduction of the shape
function.

In sec.\thinspace 3 we present an improved formula for the photon spectrum
in the rare decay 
\begin{equation}
B\rightarrow X_{s}+\gamma  \label{rare}
\end{equation}
in the threshold region. We also explicitly show that the same function, the
effective form factor $f\left( z\right) ,$ factorizes all the long-distance
effects in processes (\ref{sldec}) and (\ref{rare}).

The distributions in (\ref{sldec}) and (\ref{rare}) can be roughly divided
into two classes. The first set contains distributions \textit{not}
involving integration over the hadronic energy. A first example is the $z$
distribution in (\ref{sldec}): 
\begin{equation}
\frac{1}{\Gamma _{0}}\frac{d\Gamma _{_{SL}}}{dz}=\delta \left( 1-z\right) -%
\frac{\alpha _{S}C_{F}}{\pi }\,\mathcal{D}_{1}\left( z\right) \,-\frac{7}{4}%
\frac{\alpha _{S}C_{F}}{\pi }\,\mathcal{D}_{0}\left( z\right) +\cdots ,
\label{zdistrib}
\end{equation}
where $\Gamma _{0}=G_{F}^{2}\,m_{b}^{5}|V_{ub}|^{2}/\left( 192\pi
^{3}\right) $ is the total semileptonic width in Born approximation. A
second example is the photon spectrum in the rare decay: 
\begin{equation}
\frac{1}{\Gamma _{_{RD}}}\frac{d\Gamma _{_{RD}}}{dx_{\gamma }}=\delta \left(
1-x_{\gamma }\right) -\frac{\alpha _{S}C_{F}}{\pi }\mathcal{D}_{1}\left(
x_{\gamma }\right) \,-\frac{7}{4}\frac{\alpha _{S}C_{F}}{\pi }\,\mathcal{D}%
_{0}\left( x_{\gamma }\right) +\cdots ,  \label{xdistrib}
\end{equation}
where $\Gamma _{_{RD}}$ is the total $b\rightarrow s\gamma $ width and 
\begin{equation}
x_{\gamma }\equiv \frac{2E_{\gamma }}{m_{B}}\qquad \left( 0\leq x_{\gamma
}\leq 1\right) .
\end{equation}
In the decay (\ref{rare}), the hadronic energy is never integrated because
kinematics fixes $Q\approx m_{B}$ (see sec. 3 for a proof). As expected, the
leading terms, $\mathcal{D}_{1}\left( z\right) \,$and $\mathcal{D}_{1}\left(
x_{\gamma }\right) ,$ have the same coefficient. It is instead non trivial
that the subleading terms also, $\mathcal{D}_{0}\left( z\right) \,$and $%
\mathcal{D}_{0}\left( x_{\gamma }\right) ,$ have the same coefficient, $%
-7/4, $ and that these distributions have a perturbative expansion similar
to the one for $f\left( z\right) $ (cf. eq.\thinspace (\ref{perturb})). In
general, by measuring distributions in this class, one can directly
determine the effective form factor $f\left( z\right) ,$ or equivalently,
the shape function $\varphi \left( k_{+}\right) .$

The second class contains distributions in which the hadronic energy is
integrated over. As examples, consider the hadron-mass distribution \cite
{limetto}\ in (\ref{sldec}) 
\begin{equation}
\frac{1}{\Gamma _{0}}\frac{d\Gamma _{_{SL}}}{du}=\delta \left( 1-u\right) -%
\frac{\alpha _{S}C_{F}}{\pi }\,\mathcal{D}_{1}\left( u\right) \,-\frac{31}{12%
}\frac{\alpha _{S}C_{F}}{\pi }\,\,\mathcal{D}_{0}\left( u\right) +\cdots ,
\label{mxmb}
\end{equation}
or the electron spectrum \cite{kuhn} in (\ref{sldec}), 
\begin{equation}
\frac{1}{2\Gamma _{0}}\frac{d\Gamma _{_{SL}}}{dx_{e}}=1-\frac{\alpha
_{S}C_{F}}{2\pi }\log ^{2}\left( 1-x_{e}\right) \,-\frac{31}{12}\frac{\alpha
_{S}C_{F}}{\pi }\log \left( 1-x_{e}\right) +\cdots ,  \label{elettro}
\end{equation}
where 
\begin{equation}
x_{e}\equiv \frac{2E_{e}}{m_{B}}\qquad \left( 0\leq x_{e}\leq 1\right) .
\end{equation}
The distributions (\ref{mxmb}) and (\ref{elettro}) have the same leading
behaviour as in (\ref{zdistrib}) and (\ref{xdistrib}); the coefficient of
the subleading terms, $-31/12,$ is instead different because the integration
over the hadronic energy affects them \cite{sumrule}. The distributions in
this class are not directly related to the shape function and the extraction
of the latter from the experimental data requires in general a deconvolution.

Finally, in sec.\thinspace 4 we present our conclusions and an outlook of
future developments.

\section{Semileptonic decay}

Let us consider the hadronic tensor containing $all$ QCD dynamics: 
\begin{equation}
W_{\mu \nu }\ \equiv \sum_{X_{u}}\,\langle \,B|J_{\nu }^{+}|X_{u}\rangle
\,\langle X_{u}|J_{\mu }|B\rangle \,\,\left( 2\pi \right) ^{3}\delta
^{4}(p_{B}-q-p_{X}),  \label{Wmunu}
\end{equation}
where $q$ is the momentum of the lepton--neutrino pair and $J_{\mu }\left(
x\right) =\overline{u}\left( x\right) \gamma _{\mu }\left( 1-\gamma
_{5}\right) b\left( x\right) $ is the $b\rightarrow u$ current of the
Standard Model. The latter involves five independent form factors. We find
it convenient to use a modified parametrization with respect to the one
proposed in \cite{ndf}: 
\begin{eqnarray}
W_{\mu \nu }(p_{B};p_{X}) &=&\frac{1}{2v\cdot p_{X}}\left[ \left( n_{\mu
}v_{\nu }+n_{\nu }v_{\mu }-g_{\mu \nu }\,v\cdot n-i\epsilon _{\mu \nu \alpha
\beta }n^{\alpha }v^{\beta }\right) W_{1}\left( \varsigma ,w\right) -g_{\mu
\nu }\,W_{2}\left( \varsigma ,w\right) +\right.  \label{allabase} \\
&&\qquad \qquad \qquad \left. +v_{\mu }v_{\nu }\,W_{3}\left( \varsigma
,w\right) +\left( n_{\mu }v_{\nu }+n_{\nu }v_{\mu }\right) W_{4}\left(
\varsigma ,w\right) +n_{\mu }n_{\nu }W_{5}\left( \varsigma ,w\right) \right]
,  \notag
\end{eqnarray}
where 
\begin{equation}
v^{\mu }\equiv \frac{p_{B}^{\mu }}{m_{B}}=\left( 1;0,0,0\right)
\end{equation}
is the velocity of the beauty meson, which we take at rest, while 
\begin{equation}
n^{\mu }\equiv \frac{p_{X}^{\mu }}{v\cdot p_{X}}=\left( 1;0,0,-\sqrt{%
\varsigma }\right)
\end{equation}
is the normalized momentum of the jet, which we have taken in the minus
direction. We have defined: 
\begin{equation}
\zeta \equiv 1-\frac{m_{X}^{2}}{E_{X}^{2}}\qquad \qquad \left( 0\leq \zeta
\leq 1\right) .
\end{equation}
Note that $1-\varsigma =4\left( 1-z\right) .$ We have inserted for
convenience a factor $1/(2v\cdot p_{X})$ multiplying all the form factors.
Note that $n\cdot v=1$ and that $n^{2}=1-\varsigma \ll 1$ for $\varsigma
\lesssim 1,$ i.e. $n$ is close to the light cone in the threshold region.

The form factors can be decomposed as: 
\begin{equation}
W_{i}(z;w;\alpha _{S})=\delta _{i1}\,\,\delta (1-z)\,\theta \left(
1-w\right) +\delta _{i1}\frac{\alpha _{S}C_{F}}{\pi }\,s\left( z\right) \,+%
\frac{\alpha _{S}C_{F}}{\pi }\delta (1-z)v_{i}\left( w\right) +\frac{\alpha
_{S}C_{F}}{\pi }\,r_{i}\left( \zeta ;w\right) +O(\alpha _{S}^{2}).
\label{siparte}
\end{equation}
The function $s\left( z\right) $ contains the plus distributions, i.e. the
long-distance contributions: 
\begin{equation}
s\left( z\right) =-\mathcal{D}_{1}\left( z\right) -\frac{7}{4}\mathcal{D}%
_{0}\left( z\right) .
\end{equation}
The functions $v_{i}\left( w\right) $ come from virtual effects --- they are
proportional to $\delta (1-z)$ --- while the functions $r_{i}\left( \zeta
;w\right) $ originate from real emission only. The explicit expressions for
these functions can be extracted from \cite{ndf} and are reported in the
following. The ``virtual'' functions read: 
\begin{eqnarray}
\qquad \qquad \qquad \qquad \qquad v_{1}\left( w\right) &=&-\frac{3}{2}\log
w-\mathrm{Li}_{2}\left( 1-w\right) -\frac{w\log w}{2\left( 1-w\right) }-%
\frac{5}{4}-\frac{\pi ^{2}}{3};  \notag \\
v_{2}\left( w\right) &=&0;  \notag \\
v_{3}\left( w\right) &=&0;  \notag \\
v_{4}\left( w\right) &=&\frac{1}{2\left( 1-w\right) }\left( \frac{w\log w}{%
1-w}+1\right) ;  \notag \\
v_{5}\left( w\right) &=&\frac{w}{2\left( 1-w\right) }\left( \frac{1-2w}{1-w}%
\log w-1\right) ,
\end{eqnarray}
where $\mathrm{Li}_{2}\left( z\right) \equiv \sum_{n=1}^{\infty
}z^{n}/n^{2}\,\,\,\left( |z|\leq 1\right) $ is the standard dilogarithm. The
``real'' functions are: 
\begin{eqnarray}
r_{1}\left( \zeta ;w\right) &=&\frac{w^{2}}{4}-\frac{\left( 8-w\right)
\left( 2-w\right) }{4\zeta }+\left[ \frac{w\left( 2-w\right) }{8}+\frac{%
\left( 8-w\right) \left( 2-w\right) }{8\varsigma }\right] H\left( \varsigma
\right) +\frac{1}{1-z}\left[ H\left( \varsigma \right) +\log \left(
1-z\right) \right] ;  \notag \\
r_{2}\left( \varsigma ,w\right) &=&\frac{w\left( 8-w\right) }{8}+\frac{%
32-8w+w^{2}}{8\varsigma }-\frac{H\left( \varsigma \right) }{16}\left[
w^{2}\zeta +2w\left( 4-w\right) +\frac{32-8w+w^{2}}{\varsigma }\right] ; 
\notag \\
r_{3}\left( \varsigma ,w\right) &=&-\frac{w\left( 8-3w\right) }{8}+\frac{%
32+22w-3w^{2}}{4\varsigma }-\frac{3w\left( 12-w\right) }{8\varsigma ^{2}}+ 
\notag \\
&&+\frac{H\left( \varsigma \right) }{16}\left[ w^{2}\zeta +5w\left(
4-w\right) -\frac{64+56w-7w^{2}}{\varsigma }+\frac{3w\left( 12-w\right) }{%
\varsigma ^{2}}\right] ;  \notag \\
r_{4}\left( \varsigma ,w\right) &=&-\frac{w^{2}}{4}-\frac{w\left(
32-5w\right) }{8\varsigma }+\frac{3w\left( 12-w\right) }{8\varsigma ^{2}}-%
\frac{wH\left( \varsigma \right) }{16}\left[ 8-3w-\frac{2\left( 22-3w\right) 
}{\varsigma }+\frac{3\left( 12-w\right) }{\varsigma ^{2}}\right] ;  \notag \\
r_{5}\left( \varsigma ,w\right) &=&-\frac{w\left( 8+w\right) }{8\varsigma }-%
\frac{3w\left( 12-w\right) }{8\varsigma ^{2}}+\frac{wH\left( \varsigma
\right) }{16}\left[ w-\frac{2\left( 2-w\right) }{\varsigma }+\frac{3\left(
12-w\right) }{\varsigma ^{2}}\right] .
\end{eqnarray}
We have defined: 
\begin{equation}
H\left( \varsigma \right) \equiv \frac{1}{\sqrt{\zeta }}\log \frac{1+\sqrt{%
\zeta }}{1-\sqrt{\zeta }}.
\end{equation}
In the semi-inclusive region: 
\begin{equation}
H\left( \varsigma \right) =-\log \left( 1-z\right) +O\left( 1-z\right) ,
\end{equation}
implying that there is no $1/\left( 1-z\right) $ singularity in $r_{1}$ for $%
z\rightarrow 1.$ The functions $r_{i}$ have at most a $\log \left(
1-z\right) $ singularity for $z\rightarrow 1$, which gives no logarithmic
enhancement after integration over $z.$ Another property is that the
``real'' functions do not have any singularity of the form $1/\zeta ^{2}$
for $\zeta \rightarrow 0,$ \ but at most a singularity of the form $1/\zeta $
(the point $\varsigma =0$ corresponds to the final hadronic system at rest).
As already noted, the Born term in eq.\thinspace (\ref{siparte}) has the
additional restriction $w\leq 1.$

\subsection{Heavy mass effects}

Let us now consider the limit of infinite mass for the beauty quark, keeping
the other kinematical invariants fixed, i.e. the limit (\ref{limite1}). In
terms of our variables, this is: 
\begin{equation}
w\rightarrow 0,\quad \zeta \rightarrow \mathrm{const.}  \label{spesso}
\end{equation}
In the limit (\ref{spesso}) the ``virtual'' functions behave as: 
\begin{eqnarray}
v_{1} &\rightarrow &-\frac{3}{2}\log w;  \notag \\
\qquad \qquad \qquad \qquad \qquad \qquad \qquad v_{2} &\rightarrow &0; 
\notag \\
v_{3} &\rightarrow &0;  \notag \\
v_{4} &\rightarrow &\frac{1}{2};  \notag \\
v_{5} &\rightarrow &0.
\end{eqnarray}
The ``real'' functions have limits: 
\begin{eqnarray}
\qquad \qquad \qquad \qquad \quad \qquad r_{1} &\rightarrow &-\frac{4}{\zeta 
}+\frac{2}{\zeta }H\left( \varsigma \right) +\frac{1}{1-z}\left[ H\left(
\varsigma \right) +\log \left( 1-z\right) \right] ;\qquad \qquad \qquad
\qquad \qquad  \notag \\
r_{2} &\rightarrow &\frac{4}{\zeta }-\frac{2}{\zeta }H\left( \varsigma
\right) ;  \notag \\
r_{3} &\rightarrow &\frac{8}{\zeta }-\frac{4}{\zeta }H\left( \varsigma
\right) ;  \notag \\
r_{4} &\rightarrow &0;  \notag \\
r_{5} &\rightarrow &0.
\end{eqnarray}
Only the function $v_{1}$ diverges logarithmically in this limit: all the
other functions have a finite limit or vanish. The ``real'' functions are
finite because real emission diagrams do not generate logarithms of the
heavy mass, so $\log w$ does not appear. The logarithmic divergence of $%
v_{1} $ is associated to the term in $W_{1}:$%
\begin{equation}
W_{1}=\delta (1-z)\left[ 1-\frac{3}{2}\frac{\alpha _{S}C_{F}}{\pi }\,\log w%
\right] +\cdots .  \label{labellato}
\end{equation}
Equation (\ref{labellato}) can be understood by considering the general
properties of the vector and axial current containing a heavy and a light
field. The matrix element of these current between quark states in QCD are
of the form \cite{aggiunta}: 
\begin{equation}
\langle u|V_{\mu }|b\rangle =\overline{u}_{u}\left\{ \left[ 1+\frac{\alpha
_{S}C_{F}}{\pi }\left( \frac{3}{4}\log m_{B}+\cdots \right) \right] \gamma
_{\mu }+\frac{\alpha _{S}C_{F}}{2\pi }v_{\mu }\right\} u_{b}  \label{vettore}
\end{equation}
and 
\begin{equation}
\langle u|A_{\mu }|b\rangle =\overline{u}_{u}\left\{ \left[ 1+\frac{\alpha
_{S}C_{F}}{\pi }\left( \frac{3}{4}\log m_{B}+\cdots \right) \right] \gamma
_{\mu }\gamma _{5}-\frac{\alpha _{S}C_{F}}{2\pi }v_{\mu }\gamma _{5}\right\}
u_{b},  \label{assiale}
\end{equation}
where the dots denote terms dependent on the kinematics of the external
states. Substituting the curly brackets in the above equations in place of
the currents in eq.\thinspace (\ref{Wmunu}), we recover the logarithmic term
in eq.\thinspace (\ref{labellato}) for $W_{1},$ together with the
delta-function contribution in $W_{4}.$

The appearance of the logarithm of the heavy mass in the matrix elements (%
\ref{vettore}) and (\ref{assiale}) can be understood by considering the
effective theory: it is related to the non-conservation of the vector and
axial currents in the static limit 
\begin{equation}
m_{b}\rightarrow \infty .  \label{static}
\end{equation}
The currents are multiplicatively renormalized in the effective theory \cite
{dimenticato} and their matrix elements between quark states read: 
\begin{eqnarray}
\qquad \qquad \qquad \qquad \qquad \qquad \langle u|\widetilde{V}_{\nu
}\left( \mu \right) |b\rangle &=&\overline{u}_{u}\left[ 1+\frac{\alpha
_{S}C_{F}}{\pi }\left( \frac{3}{4}\log \mu +\cdots \right) \right] \gamma
_{\nu }u_{b},  \notag \\
\langle u|\widetilde{A}_{\nu }\left( \mu \right) |b\rangle &=&\overline{u}%
_{u}\left[ 1+\frac{\alpha _{S}C_{F}}{\pi }\left( \frac{3}{4}\log \mu +\cdots
\right) \right] \gamma _{\nu }\gamma _{5}u_{b},
\end{eqnarray}
where $\mu $ is the renormalization point and the dots denote $\mu $%
-independent terms. As expected, the logarithmic dependence on the
renormalization point $\mu $ of the effective currents matches the
dependence on the heavy mass $m_{b}$ in the matrix elements of the full QCD
currents.

\subsection{Infrared factorization}

To perform factorization of infrared logarithms, the form factors are
conveniently written as: 
\begin{equation}
W_{i}(z;w;\alpha _{S})=\delta _{i1}\,f\left( z;\alpha _{S}\right) \,+\frac{%
\alpha _{S}C_{F}}{\pi }\,\delta (1-z)\,v_{i}\left( w\right) +\frac{\alpha
_{S}C_{F}}{\pi }\,r_{i}\left( \zeta ;w\right) -\delta _{i1}\delta
(1-z)\theta \left( w-1\right) +O(\alpha _{S}^{2}).\quad
\end{equation}
The function $f\left( z\right) ,$ defined previously (eq.\thinspace (\ref
{perturb})), contains the long-distance effects. The soft-gluon resummation
of $f(z)$ allows us to describe the semi-inclusive region 
\begin{equation}
1-z\ll 1
\end{equation}
and has been performed to NLO in \cite{civuole,civuole2,ultimo}; $\ f(z)$ is
perturbatively computable as long as 
\begin{equation}
1-z\gg \frac{\Lambda }{Q},
\end{equation}
where $\Lambda $ is the QCD scale. In the region 
\begin{equation}
1-z\sim \frac{\Lambda }{Q}
\end{equation}
$f(z)$ acquires a substantial non-perturbative component related to the
well-known Fermi-motion effects \cite{penultimo}.

The moments of the effective form factor, 
\begin{equation}
f_{N}\equiv \int_{0}^{1}dz\,z^{N-1}\,f(z),
\end{equation}
can be written as the exponential of a series of functions: 
\begin{equation}
f_{N}=\exp \left[ L\,\,g_{1}\left( \beta _{0}\alpha _{S}L\right)
+g_{2}\left( \beta _{0}\alpha _{S}L\right) +\alpha _{S}\,g_{3}\left( \beta
_{0}\alpha _{S}L\right) +\cdots \right] ,  \label{seriefun}
\end{equation}
where the leading and next-to-leading functions have the simple analytical
expressions \cite{civuole,civuole2,ultimo}: 
\begin{eqnarray}
g_{1}\left( \lambda \right) &=&-\frac{A_{1}}{2\beta _{0}}\,\frac{1}{\lambda }%
\left[ \left( 1-2\lambda \right) \log \left( 1-2\lambda \right) -2\left(
1-\lambda \right) \log \left( 1-\lambda \right) \right] ;  \notag \\
\,g_{2}\left( \lambda \right) &=&\frac{\beta _{0}A_{2}-\beta _{1}A_{1}}{%
2\beta _{0}^{3}}\left[ \log (1-2\lambda )-2\log (1-\lambda )\right] -\frac{%
\beta _{1}A_{1}}{4\beta _{0}^{3}}\left[ \log ^{2}(1-2\lambda )-2\log
^{2}(1-\lambda )\right] +  \notag \\
&&+\frac{S_{1}}{2\beta _{0}}\log (1-2\lambda )+\frac{C_{1}}{\beta _{0}}\log
(1-\lambda ).  \label{compatta}
\end{eqnarray}
We have defined $L\equiv \log n$ and $n\equiv N/N_{0},$ with $N_{0}\equiv
e^{-\gamma _{E}}=0.561459\ldots $ and $\gamma _{E}=0.577216\ldots $ the
Euler constant. The first two coefficients of the $\beta $-function are: 
\begin{equation}
\beta _{0}=\frac{11C_{A}-2n_{F}}{12\pi }=\frac{33-2n_{F}}{12\pi },\qquad
\beta _{1}=\frac{17C_{A}^{2}-5C_{A}n_{F}-3C_{F}n_{F}}{24\pi ^{2}}=\frac{%
153-19\,n_{F}}{24\pi ^{2}},
\end{equation}
where $C_{A}=N_{c}=3$ and $n_{F}=3$ is the number of active quark flavours.
The one-loop quantities $S_{1}$ and $C_{1}$ have the values: 
\begin{equation}
S_{1}=-\frac{C_{F}}{\pi },\qquad \qquad C_{1}=-\frac{3}{4}\frac{C_{F}}{\pi }.
\end{equation}
The two-loop quantity $A_{2}$ is given by \cite{kodtren, cattren}: 
\begin{equation}
A_{2}=\frac{C_{F}}{2\pi ^{2}}K
\end{equation}
where, in the $\overline{MS}$ \ scheme for the coupling constant, 
\begin{equation}
K=C_{A}\left( \frac{67}{18}-\frac{\pi ^{2}}{6}\right) -\frac{10}{9}%
n_{f}T_{R},
\end{equation}
with $T_{R}=1/2.$

The original form factor $f\left( z\right) ,$ in momentum space, is obtained
with an inverse Mellin transform of $f_{N}$ in (\ref{seriefun}): 
\begin{equation}
f\left( z\right) =\int_{c-i\infty }^{c+i\infty }\frac{dN}{2\pi i}%
\,z^{-N}\,f_{N},  \label{inversa}
\end{equation}
where the constant $c$ is chosen so that all the singularities of $f_{N}$
lie to the left of the integration contour. The inverse transform is usually
done numerically \cite{mangano}. \ In leading order, the transform from $%
f_{N}$ to the cumulative distribution 
\begin{equation}
F\left( z\right) \equiv \int_{z}^{1}f\left( z^{\prime }\right) dz^{\prime }
\end{equation}
is equivalent to the simple replacement: 
\begin{equation}
\log n\rightarrow -\log y,
\end{equation}
where 
\begin{equation}
y\equiv 1-z=\frac{m_{X}^{2}}{Q^{2}}.
\end{equation}
The leading form factor can then be explicitly given in momentum space and
reads: 
\begin{equation}
\,f(z;\alpha _{S})_{l}=\frac{d}{dy}\exp \left[ h\left( y;\alpha _{S}\right) %
\right] ,  \label{leading}
\end{equation}
where 
\begin{equation}
h\left( y;\alpha _{S}\right) \equiv -\frac{A_{1}}{2\beta _{0}^{2}\alpha _{S}}%
\left[ \left( 1+2\beta _{0}\alpha _{S}\log y\right) \log \left( 1+2\beta
_{0}\alpha _{S}\log y\right) -2\left( 1+\beta _{0}\alpha _{S}\log y\right)
\log \left( 1+\beta _{0}\alpha _{S}\log y\right) \right] \,.  \label{defh}
\end{equation}
It holds that $h\left( y=1\right) =0.$ In agreement with the frozen coupling
case (see later) we $define$: $\exp \left[ h\left( y=0\right) \right] =0.$
The QCD coupling is evaluated at the hard scale of the process: 
\begin{equation}
\alpha _{S}\equiv \alpha _{S}\left( Q^{2}\right) =\alpha _{S}\left(
w^{2}m_{B}^{2}\right) .  \label{nuova}
\end{equation}
To some approximation, we can set $w=1$ in the running coupling so that 
\begin{equation}
\alpha _{S}\simeq \alpha _{S}\left( m_{B}^{2}\right) .
\end{equation}
Equation (\ref{leading}) is simple but non-trivial, as the presence of the
singularity for \cite{penultimo} 
\begin{equation}
y\leq y_{\min }\equiv \exp \left[ -\frac{1}{2\beta _{0}\alpha _{S}}\right]
\sim \frac{\Lambda }{Q}
\end{equation}
shows. This singularity is related to the infrared pole in the running
coupling. The frozen-coupling case is obtained by taking the limit $\beta
_{0}\rightarrow 0$ on the r.h.s of eq.\thinspace (\ref{defh}) and reads:%
\footnote{%
The plus regularization of the function $\log \left( 1-z\right) /\left(
1-z\right) $ in the last member can be omitted because the exponential
suppresses the virtual contribution for $\alpha _{S}\neq 0.$} 
\begin{eqnarray}
\qquad \qquad f(z,\alpha _{S}) &\approx &\frac{d}{dy}\exp \left[ -\frac{%
\alpha _{S}C_{F}}{2\pi }\log ^{2}y\right]  \notag \\
&=&-\frac{\alpha _{S}C_{F}}{\pi }\frac{\log y}{y}\exp \left[ -\frac{\alpha
_{S}C_{F}}{2\pi }\log ^{2}y\right] \qquad \qquad \qquad \left( \beta
_{0}=0\right) .
\end{eqnarray}

Our task is to factorize the long-distance contributions found in the
hadronic tensor, i.e. the terms containing distributions.\ The simplest
factorization \ scheme involves a minimal subtraction, so that the form
factors are written: 
\begin{equation}
W_{i}\left( \varsigma ,w;\alpha _{S}\right) =V_{i}\left( w;\alpha
_{S}\right) \,f(z;\alpha _{S})+R_{i}\left( \varsigma ,w;\alpha _{S}\right) .
\end{equation}
According to the above decomposition, the hadronic tensor is: 
\begin{equation}
W_{\mu \nu }\left( \varsigma ,w;\alpha _{S}\right) =V_{\mu \nu }\left(
w;\alpha _{S}\right) \,f(z;\alpha _{S})+R_{\mu \nu }\left( \varsigma
,w;\alpha _{S}\right) .
\end{equation}
The tensors containing the virtual effects and the real ones are defined in
a way analogous to the hadronic tensor: 
\begin{eqnarray}
V_{\mu \nu }\left( w;\alpha _{S}\right) &=&\frac{1}{2v\cdot p_{X}}\left[
\left( n_{\mu }v_{\nu }+n_{\nu }v_{\mu }-g_{\mu \nu }\,v\cdot n-i\epsilon
_{\mu \nu \alpha \beta }n^{\alpha }v^{\beta }\right) V_{1}\left( w\right)
-g_{\mu \nu }\,V_{2}\left( w\right) +\right. \\
&&\qquad \qquad \qquad \qquad \left. +v_{\mu }v_{\nu }\,V_{3}\left( w\right)
+\left( n_{\mu }v_{\nu }+n_{\nu }v_{\mu }\right) V_{4}\left( w\right)
+n_{\mu }n_{\nu }V_{5}\left( w\right) \right] ,  \notag
\end{eqnarray}
and 
\begin{eqnarray}
R_{\mu \nu }\left( \varsigma ,w;\alpha _{S}\right) &=&\frac{1}{2v\cdot p_{X}}%
\left[ \left( n_{\mu }v_{\nu }+n_{\nu }v_{\mu }-g_{\mu \nu }\,v\cdot
n-i\epsilon _{\mu \nu \alpha \beta }n^{\alpha }v^{\beta }\right) R_{1}\left(
\varsigma ,w\right) -g_{\mu \nu }\,R_{2}\left( \varsigma ,w\right) +\right.
\\
&&\qquad \qquad \qquad \qquad \left. +v_{\mu }v_{\nu }\,R_{3}\left(
\varsigma ,w\right) +\left( n_{\mu }v_{\nu }+n_{\nu }v_{\mu }\right)
R_{4}\left( \varsigma ,w\right) +n_{\mu }n_{\nu }R_{5}\left( \varsigma
,w\right) \right] .  \notag
\end{eqnarray}
The form factors have an expansion in powers of $\alpha _{S}:$ 
\begin{equation}
V_{i}\left( w;\alpha _{S}\right) =\delta _{i1}+\frac{\alpha _{S}C_{F}}{\pi }%
v_{i}\left( w\right) +\left( \frac{\alpha _{S}}{\pi }\right)
^{2}v_{i}^{\prime }\left( w\right) +\cdots
\end{equation}
and 
\begin{equation}
R_{i}\left( \varsigma ,w;\alpha _{S}\right) =-\delta _{i1}\delta \left(
1-z\right) \theta \left( w-1\right) +\frac{\alpha _{S}C_{F}}{\pi }%
r_{i}\left( \varsigma ,w\right) +\left( \frac{\alpha _{S}}{\pi }\right)
^{2}r_{i}^{\prime }\left( \varsigma ,w\right) +\cdots .
\end{equation}
All the dependence on the heavy mass, i.e. on $w,$ is contained into the
short-distance form factors: the function $f(z;\alpha _{S})$ depends only on
the ratio of the final-state kinematical variables.

In leading approximation, the hadronic tensor has the particularly simple
form: 
\begin{equation}
W_{\mu \nu }\left( z,w;\alpha _{S}\right) _{l}=\frac{1}{m_{B}\,w}\left(
n_{\mu }v_{\nu }+n_{\nu }v_{\mu }-g_{\mu \nu }\,v\cdot n-i\epsilon _{\mu \nu
\alpha \beta }n^{\alpha }v^{\beta }\right) \,\frac{d}{dy}\exp \left[ h\left(
y;\alpha _{S}\right) \right] ,
\end{equation}
with $\,h\left( y;\alpha _{S}\right) $ given by eq.\thinspace (\ref{defh}).

\subsection{Triple differential distribution}

Let us now consider the most general distribution in (\ref{sldec}), which is
a triple differential distribution. One has basically to contract the
hadronic tensor with the leptonic one. A third kinematical variable is
involved, which we choose as the electron energy: $x\equiv x_{e}.$ The
expression in terms of the form factors reads: 
\begin{equation}
\frac{1}{12\Gamma _{0}}\frac{d^{3}\Gamma }{dxdwdz}\left( z,w,x;\alpha
_{S}\right) =\sum_{i=1}^{5}P_{i}\left( x,w,z\right) \,W_{i}\left( w,z;\alpha
_{S}\right) ,
\end{equation}
where $P_{i}\left( x,w,z\right) $ are polynomials in all the kinematical
variables (independent of $\alpha _{S}$): 
\begin{eqnarray}
\qquad \qquad \qquad P_{1}\left( x,w,z\right) &=&\left[ 1+\overline{x}-w%
\right] \left[ w-\overline{x}-\left( 1-z\right) w^{2}\right] ;  \notag \\
P_{2}\left( x,w,z\right) &=&\frac{w}{2}\left[ 1-w+\left( 1-z\right) w^{2}%
\right] ;  \notag \\
P_{3}\left( x,w,z\right) &=&\frac{w}{4}\left[ \overline{x}\left( w-\overline{%
x}\right) -\left( 1-z\right) w^{2}\right] ;  \notag \\
P_{4}\left( x,w,z\right) &=&\overline{x}\left( w-\overline{x}\right) -\left(
1-z\right) w^{2};  \notag \\
P_{5}\left( x,w,z\right) &=&\frac{1}{w}\left[ \overline{x}\left( w-\overline{%
x}\right) -\left( 1-z\right) w^{2}\right] .
\end{eqnarray}
Explicitly, one has: 
\begin{eqnarray}
\frac{1}{12\Gamma _{0}}\frac{d^{3}\Gamma }{dxdwdz}\left( z,w,x;\alpha
_{S}\right) &=&\left[ 1+\overline{x}-w\right] \left[ w-\overline{x}-\left(
1-z\right) w^{2}\right] W_{1}+\frac{w}{2}\left[ 1-w+\left( 1-z\right) w^{2}%
\right] W_{2}+  \notag \\
&&+\left[ \overline{x}\left( w-\overline{x}\right) -\left( 1-z\right) w^{2}%
\right] \left[ \frac{w}{4}W_{3}+W_{4}+\frac{1}{w}W_{5}\right] ,
\end{eqnarray}
where $\overline{x}\equiv 1-x.$

Let us now consider the kinematical constraints; there are various cases 
\cite{ndf}. For a given electron energy, in the range 
\begin{equation}
0\leq \overline{x}\leq 1,
\end{equation}
the range of the hadronic energy is 
\begin{equation}
\overline{x}\leq w\leq 1+\overline{x},
\end{equation}
and the range of $y$ is: 
\begin{equation}
\max \left( 0,\frac{w-1}{w^{2}}\right) \leq y\leq \frac{\overline{x}\left( w-%
\overline{x}\right) }{w^{2}}.  \label{elei}
\end{equation}
For a given hadronic energy $w$ in the range 
\begin{equation}
0\leq w\leq 2,
\end{equation}
the range of $y$ is: 
\begin{equation}
\qquad \max \left[ 0,\frac{w-1}{w^{2}}\right] \leq y,  \label{nonsup}
\end{equation}
and the range of the electron energy is: 
\begin{equation}
\frac{w}{2}\left( 1-\sqrt{\varsigma }\right) \leq \overline{x}\leq \frac{w}{2%
}\left( 1+\sqrt{\varsigma }\right) .  \label{xallafine}
\end{equation}
For a given hadronic mass $\zeta $ in the range 
\begin{equation}
0\leq \varsigma \leq 1,
\end{equation}
the range of the hadronic energy is: 
\begin{equation}
w\leq \frac{2}{1+\sqrt{\varsigma }}  \label{piudura}
\end{equation}
and the range of the electron energy is again given by eq.\thinspace (\ref
{xallafine}).

In the semi-inclusive region, the triple differential distribution is
naturally written as: 
\begin{equation}
\frac{1}{12\Gamma _{0}}\frac{d^{3}\Gamma }{dxdwdz}\left( z,w,x;\alpha
_{S}\right) =C\left( w,x;\alpha _{S}\right) \,f\left( z;\alpha _{S}\right)
+D\left( \varsigma ,w,x;\alpha _{S}\right) .  \label{bellissima}
\end{equation}
The coefficient function and the remainder function have a power series
expansion in $\alpha _{S}:$%
\begin{eqnarray}
C\left( w,x;\alpha _{S}\right) &=&c_{0}\left( w,x\right) +\frac{\alpha
_{S}C_{F}}{\pi }\,c\left( w,x\right) +\left( \frac{\alpha _{S}}{\pi }\right)
^{2}\,c^{\prime }\left( w,x\right) +\cdots  \notag \\
D\left( z,w,x;\alpha _{S}\right) &=&\frac{\alpha _{S}C_{F}}{\pi }\,d\left(
z,w,x\right) +\left( \frac{\alpha _{S}}{\pi }\right) ^{2}d^{\prime }\left(
z,w,x\right) +\cdots .
\end{eqnarray}
The function $D$ has at most a logarithmic singularity $\sim \log ^{k}\left(
1-z\right) $ for $z\rightarrow 1$ and its lowest order vanishes: $%
d_{0}\left( z,w,x\right) =0.$ The explicit expression of the coefficient
function reads: 
\begin{equation}
C\left( w,x;\alpha _{S}\right) =\sum_{i=1}^{5}P_{i}\left( x,w,1\right)
\,V_{i}\left( w;\alpha _{S}\right) =P_{1}\left( x,w,1\right) +\frac{\alpha
_{S}C_{F}}{\pi }\sum_{i=1}^{5}P_{i}\left( x,w,1\right) \,v_{i}\left(
w\right) +\cdots .
\end{equation}
Expanding in powers of \ $\alpha _{S}$ on both sides, the first two terms of
the coefficient function read: 
\begin{eqnarray}
c_{0}\left( w,x\right) &=&P_{1}\left( x,w,1\right) =\left( 1+\overline{x}%
-w\right) \left( w-\overline{x}\right) ;  \label{czero} \\
c\left( w,x\right) &=&\sum_{i=1}^{5}P_{i}\left( x,w,1\right) \,v_{i}\left(
w\right) \\
&=&\left( 1+\overline{x}-w\right) \left( w-\overline{x}\right) \left[ -\frac{%
3}{2}\log w-\mathrm{Li}_{2}\left( 1-w\right) -\frac{w\log w}{2\left(
1-w\right) }-\frac{5}{4}-\frac{\pi ^{2}}{3}\right] +\overline{x}\left( w-%
\overline{x}\right) \frac{\log w}{2\left( 1-w\right) }.  \notag
\end{eqnarray}
The remainder function reads: 
\begin{equation}
D\left( z,w,x;\alpha _{S}\right) =\sum_{i=1}^{5}P_{i}\left( x,w,z\right)
\,R_{i}\left( w,z;\alpha _{S}\right) +\sum_{i=1}^{5}\left[ P_{i}\left(
x,w,z\right) -P_{i}\left( x,w,1\right) \right] \,V_{i}\left( w;\alpha
_{S}\right) \,f\left( z;\alpha _{S}\right) .  \label{resto}
\end{equation}
Since 
\begin{equation}
P_{i}\left( x,w,z\right) -P_{i}\left( x,w,1\right) =O\left( 1-z\right) ,
\end{equation}
one can replace the fixed-order expansion of $f\left( z;\alpha _{S}\right) $
in the r.h.s. of \ eq.\thinspace (\ref{resto}) and neglect the virtual
contributions, i.e. the plus regularization.\ The one-loop contribution to
the remainder function is: 
\begin{equation}
d\left( z,w,x\right) =\sum_{i=1}^{5}P_{i}\left( x,w,z\right) \,r_{i}\left(
w,z\right) +\left[ P_{1}\left( x,w,1\right) -P_{1}\left( x,w,z\right) \right]
\frac{\log \left( 1-z\right) +7/4}{1-z}.
\end{equation}
The explicit expression for $d$ is quite long and we do not report it here.

Equation (\ref{bellissima}) is our main result and allows computing an
arbitrary distribution in the threshold region to NLO logarithmic accuracy.
It is the generalization to a triple differential distribution of the
representation for the resummed shape variables for very small values of the
resolution parameters, such as the thrust distribution for $1-T\ll 1$ \cite
{thrust}.

In leading order, one needs the effective form factor $\,f\left( z;\alpha
_{S}\right) $ in double-logarithmic approximation, eq.\thinspace (\ref
{leading}), and the coefficient function at the Born level given above, $%
c_{0}\left( w,x\right) ,$ in eq.\thinspace (\ref{czero}). Replacing the
expressions for the coefficient function and the effective form factor, the
distribution reads, in leading order: 
\begin{eqnarray}
\frac{1}{\Gamma _{0}}\frac{d^{3}\Gamma }{dxdwdy} &=&12\left( 1+\overline{x}%
-w\right) \left( w-\overline{x}\right) \,f\left( z;\alpha _{S}\right)  \notag
\\
&=&\,12\left( 1+\overline{x}-w\right) \left( w-\overline{x}\right) \frac{d}{%
dy}\exp \left[ h\left( y;\alpha _{S}\right) \right] ,  \label{tripla}
\end{eqnarray}
with $h\left( y\right) $ given in eq.\thinspace (\ref{defh}).

In next-to-leading order (NLO), one needs $\,f\left( z;\alpha _{S}\right) $
to single-logarithmic accuracy and the one-loop functions $c\left(
w,x\right) $ and $d\left( z,w,x\right) .$ Since the term containing the
long-distance effects depends only on $z,$ the integrations over the
electron energy and the hadronic energy do not touch the infrared
logarithmic structure.

\subsection{Distribution in the hadron and electron energy}

Integrating over the variable $y$ on both sides of eq.\thinspace (\ref
{tripla}), in the range (\ref{elei}), one obtains for the leading
distribution in the energies $w$ and $x$: 
\begin{equation}
\frac{1}{\Gamma _{0}}\frac{d^{2}\Gamma }{dxdw}=12\left( 1+\overline{x}%
-w\right) \left( w-\overline{x}\right) \left\{ \exp \left[ h\left( \frac{%
\overline{x}\left( w-\overline{x}\right) }{w^{2}}\right) \right] -\theta
\left( \Delta w\right) \exp \left[ h\left( \frac{\Delta w}{w^{2}}\right) %
\right] \right\} ,
\end{equation}
where 
\begin{equation}
\Delta w\equiv w-1.
\end{equation}
Since in the term proportional to $\theta \left( \Delta w\right) ,$ the
hadron energy is restricted to the range $1\leq w\leq 2,$ according to the
leading logarithmic approximation, we can set $w^{2}=1.$ Introducing the
neutrino energy $x_{\nu }\equiv 2E_{\nu }/m_{B}$ in place of the hadron
energy, satisfying 
\begin{equation}
x_{e}+x_{\nu }+w=2,
\end{equation}
the distribution can be written in the more ``symmetrical'' form: 
\begin{equation}
\frac{1}{\Gamma _{0}}\frac{d^{2}\Gamma }{dx_{e}dx_{\nu }}\simeq 12x_{\nu
}\left( 1-x_{\nu }\right) \left\{ \exp \left[ h\left( \frac{\left(
1-x_{e}\right) \left( 1-x_{\nu }\right) }{\left( 2-x_{e}-x_{\nu }\right) ^{2}%
}\right) \right] \,-\,\theta \left( 1-x_{e}-x_{\nu }\right) \exp \left[
h\left( 1-x_{e}-x_{\nu }\right) \right] \right\} .
\end{equation}

\psfig{bbllx=90pt, bblly=335pt, bburx=650pt, bbury=740pt,
file=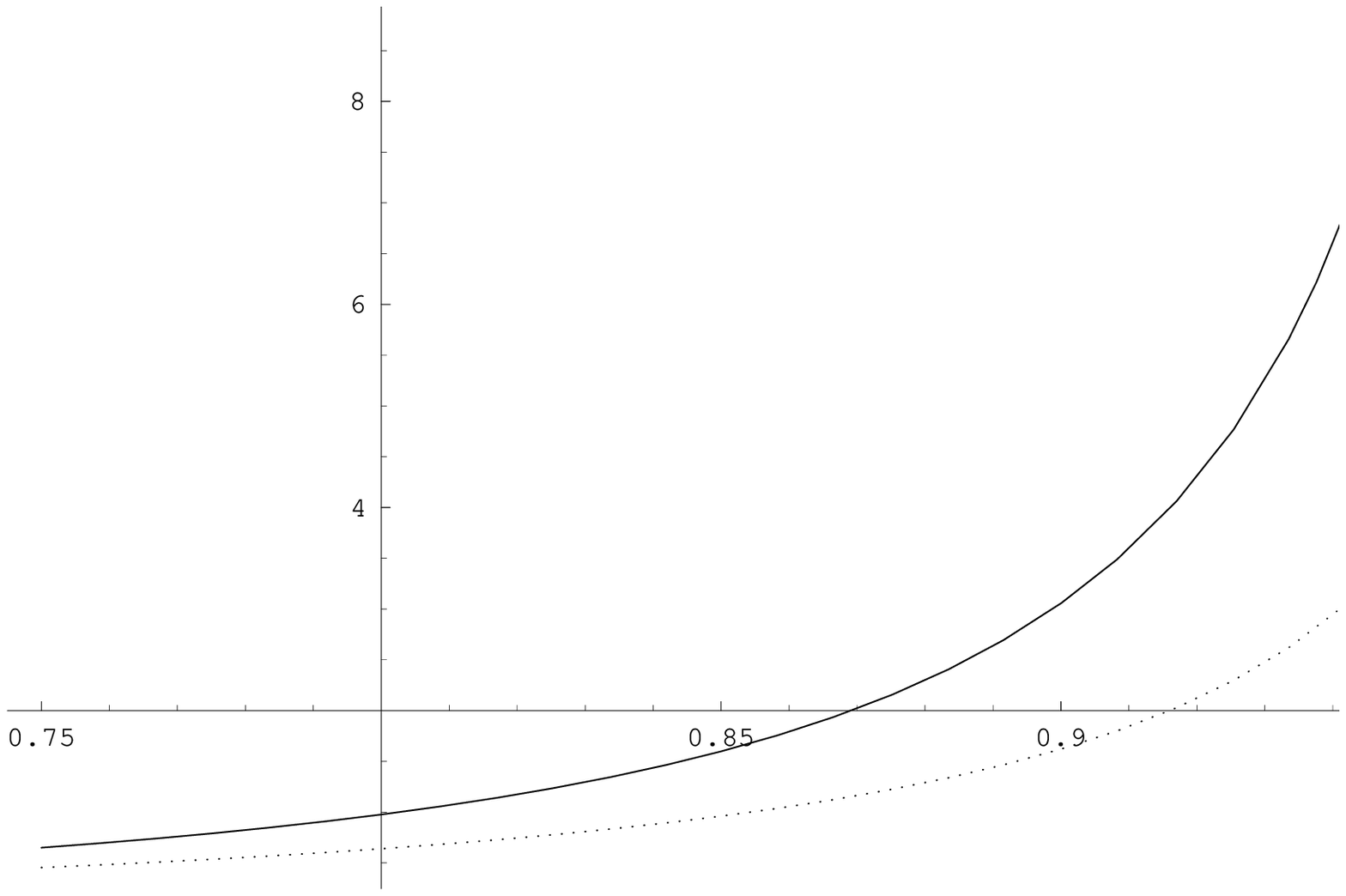, height=9cm, width=13cm} \vspace{-1.5cm}

\noindent \textit{Fig.~1: Plot of the }$z$\textit{\ distribution in leading
logarithmic approximation in the range }$0.75\leq z\leq 0.95.$\textit{\
Solid line: running--coupling case; dotted line: frozen--coupling case.}

\noindent Since we are interested to the semi-inclusive region $1-x_{e}\ll 1$%
, we can set $x_{e}=1$ whenever possible, so that: 
\begin{equation}
\frac{1}{\Gamma _{0}}\frac{d^{2}\Gamma }{dx_{e}dx_{\nu }}\simeq 12x_{\nu
}\left( 1-x_{\nu }\right) \left\{ \exp \left[ h\left( \frac{1-x_{e}}{%
1-x_{\nu }}\right) \right] \,-\,\theta \left( 1-x_{e}-x_{\nu }\right) \exp %
\left[ h\left( 1-x_{e}-x_{\nu }\right) \right] \right\} .
\end{equation}
In the frozen-coupling limit, the above expression reduces to: 
\begin{equation}
\frac{1}{\Gamma _{0}}\frac{d^{2}\Gamma }{dx_{e}dx_{\nu }}\simeq 12x_{\nu
}\left( 1-x_{\nu }\right) \left\{ \exp \left[ -\frac{\alpha _{S}C_{F}}{2\pi }%
\log ^{2}\left( \frac{1-x_{e}}{1-x_{\nu }}\right) \right] -\theta \left(
1-x_{e}-x_{\nu }\right) \exp \left[ -\frac{\alpha _{S}C_{F}}{2\pi }\log
^{2}\left( 1-x_{e}-x_{\nu }\right) \right] \right\} .
\end{equation}
The expansion to first order in $\alpha _{S}$ reads: 
\begin{equation}
\frac{1}{\Gamma _{0}}\frac{d^{2}\Gamma }{dx_{e}dx_{\nu }}\approx 12x_{\nu
}\left( 1-x_{\nu }\right) \left\{ \theta \left( x_{e}+x_{\nu }-1\right) -%
\frac{\alpha _{S}C_{F}}{2\pi }\log ^{2}\left( \frac{1-x_{e}}{1-x_{\nu }}%
\right) +\theta \left( 1-x_{e}-x_{\nu }\right) \frac{\alpha _{S}C_{F}}{2\pi }%
\log ^{2}\left( 1-x_{e}-x_{\nu }\right) \right\} .
\end{equation}
This formula contains the same double logarithms as the one-loop
distribution computed in \cite{ndf}.

\subsection{Distribution in the hadronic variables $z$ and $w$}

Performing the integration over the electron energy in the range (\ref
{xallafine}), the double distribution in $z$ and $w$ reads, to leading
logarithmic accuracy: 
\begin{equation}
\frac{1}{\Gamma _{0}}\frac{d^{2}\Gamma }{dwdy}=2w^{2}\left( 3-2w\right)
\,f\left( y\right) =2w^{2}\left( 3-2w\right) \,\frac{d}{dy}\exp \left[
h\left( y;\alpha _{S}\right) \right] .  \label{leadwz}
\end{equation}
Note the complete factorization in the variables $w$ and $y.$ Measuring this
distribution, it is possible to determine in a direct way the shape function
and to check the factorized structure.

\psfig{bbllx=90pt, bblly=335pt, bburx=650pt, bbury=740pt,
file=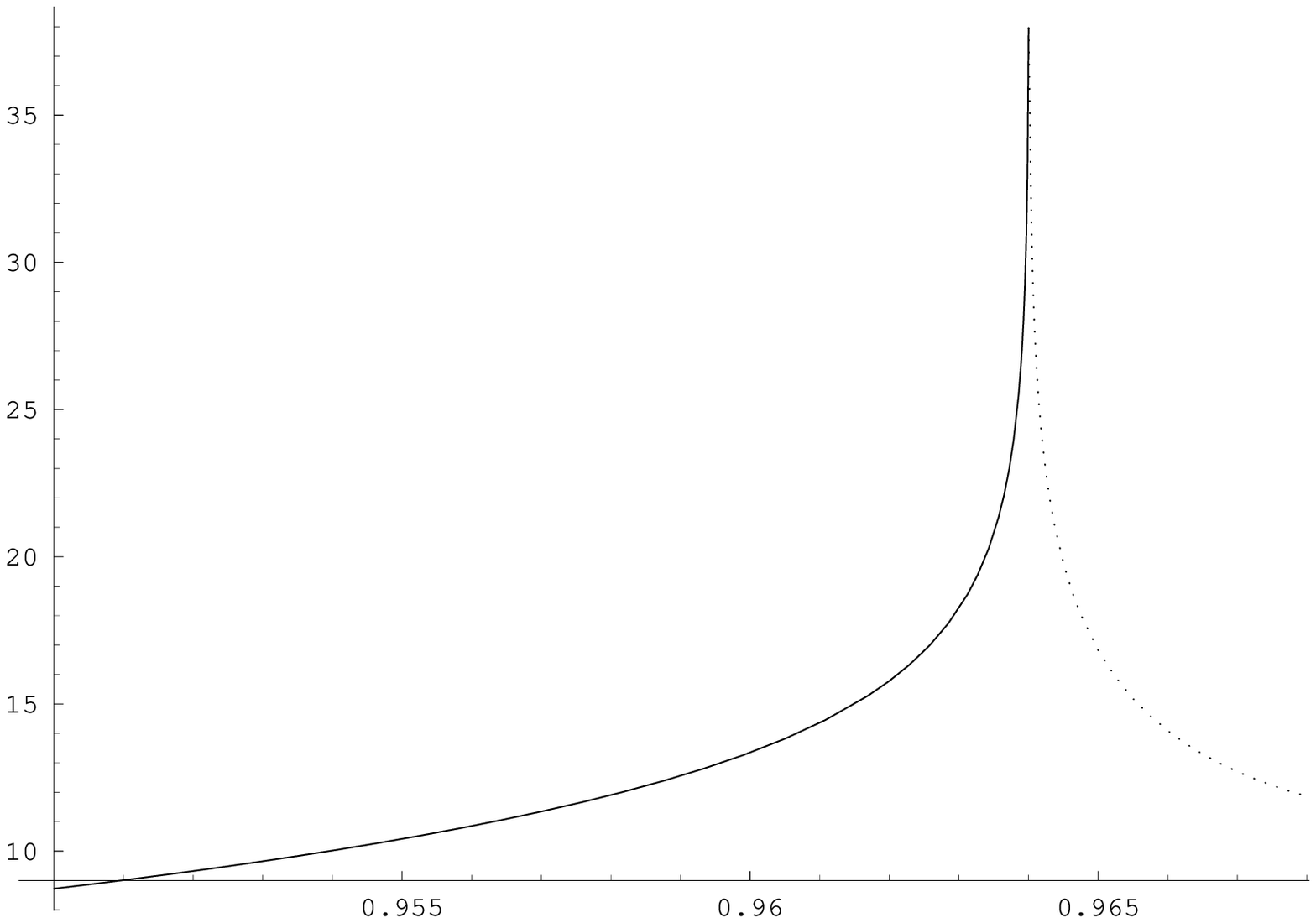, height=9cm, width=13cm} \vspace{-1.5cm}

\noindent \textit{Fig.~2: Plot of the }$z$\textit{\ distribution in the
running coupling case for }$z\geq 0.95,$\textit{\ i.e. for very large }$z.$%
\textit{\ The dotted line represents the real part of the distribution, as
the latter acquires an imaginary part after the peak.}

\subsection{Distribution in $z$}

Integrating the above distribution over the hadronic energy in the range (%
\ref{piudura}), 
\begin{equation}
w\leq 1+O\left( y\right) ,
\end{equation}
one obtains the leading distribution in the (square of the) hadron mass
normalized to the hadron energy: 
\begin{equation}
\frac{1}{\Gamma _{0}}\frac{d\Gamma }{dy}=f\left( y\right) =\frac{d}{dy}\exp %
\left[ h\left( y\right) \right] .  \label{inz}
\end{equation}
Note that this distribution does not coincide, beyond the double logarithm
at one loop, with the hadron mass distribution, as discussed in the
introduction (cf. eq.\thinspace (\ref{mxmb})). Since $f\left( y\right) $ is
proportional to the shape function via a short-distance factor, a measure of
this distribution allows a direct determination of the shape function. The
distribution (\ref{inz}) is plotted in fig. 1. The region very close to $z=1$
is plotted in fig.\thinspace 2. The maximum of $f\left( z\right) $ occurs at 
\begin{equation}
z_{\max }\sim 1-\frac{\Lambda }{Q},
\end{equation}
i.e. in the non-perturbative region described by the shape function\footnote{%
I wish to thank S. Catani for a discussion on this point.}. The leading
distribution (\ref{inz}) acquires indeed an imaginary part for $z>z_{\max }.$
\ 

\subsection{Hadron energy spectrum}

Integrating over $y$ the distribution (\ref{leadwz}) according to condition (%
\ref{nonsup}) (the upper limit on $y$ can be taken to be one for
simplicity's sake), one obtains the resummed hadron energy spectrum in
leading approximation: 
\begin{equation}
\frac{1}{\Gamma _{0}}\frac{d\Gamma }{dw}=2w^{2}\left( 3-2w\right) \left\{
1-\theta \left( w-1\right) \,\,\exp \left[ h\left( w-1\right) \right]
\right\} .
\end{equation}
The above distribution acquires an imaginary part --- and therefore is
completely unphysical --- in the region 
\begin{equation}
0\leq \Delta w\leq \exp \left[ -\frac{1}{2\beta _{0}\,\alpha _{S}\left(
m_{B}^{2}\right) }\right] \sim \frac{\Lambda }{m_{B}}.
\end{equation}
This effect is related to the Landau pole in the coupling and implies that
the region 
\begin{equation}
E_{X}\,\in \,\left[ \frac{m_{B}}{2},\,\frac{m_{B}}{2}+O\left( \Lambda
\right) \right]
\end{equation}
is non-perturbative. This region is described by the shape function because
it is related to the integration over $y$ down to $\Delta w\ll 1.$

In the frozen-coupling case, one has the simple expression: 
\begin{equation}
\frac{1}{\Gamma _{0}}\frac{d\Gamma }{dw}\approx 2w^{2}\left( 3-2w\right)
\left\{ 1-\theta \left( w-1\right) \,\,\exp \left[ -\frac{\alpha _{S}C_{F}}{%
2\pi }\log ^{2}\left( w-1\right) \right] \right\} \qquad \qquad \left( \beta
_{0}=0\right) .  \label{congelato}
\end{equation}
The expansion to order $\alpha _{S}$ of the r.h.s. is in agreement with the
double logarithmic approximation of the one-loop distribution computed in 
\cite{ndf}: 
\begin{equation}
\frac{1}{\Gamma _{0}}\frac{d\Gamma }{dw}=2w^{2}\left( 3-2w\right) \left\{
\theta \left( 1-w\right) +\theta \left( w-1\right) \,\,\left[ \frac{1}{2}%
\frac{\alpha _{S}C_{F}}{\pi }\log ^{2}\left( w-1\right) +\frac{7}{4}\frac{%
\alpha _{S}C_{F}}{\pi }\log \left( w-1\right) \right] +\cdots \right\} .
\end{equation}
Note the factor $7/4$ in front of the single logarithm, which is
characteristic of the distributions not integrated over the hadron energy,
as discussed in the introduction. The hadron spectrum is plotted in fig. 3.
Let us remark that, in the frozen-coupling case, the distribution is smooth
in the point $w=1,$ because the exponential in eq.\thinspace (\ref{congelato}%
) goes to zero together with all its derivatives for $w\rightarrow +1.$ A
plot of a small neighborhood of this point would indeed show that the
apperent cusp in fig.\thinspace 3 is actually not there.  

\section{Rare decay}

In this section we present an improved expression for the photon spectrum
near the endpoint in the rare decay (\ref{rare}) \cite{citare,inostri,misiak}%
. The only terms in the spectrum containing large infrared logarithms to $%
O\left( \alpha _{S}\right) $ involve two insertions of the operator $O_{7}$
(see \cite{misiak} for the definition of the operator basis). The latter is
also the dominant one in the inclusive rate, so this operator usually is the
only one considered. It is however possible to improve the distribution by
including the other operators, as we are going to show.

The hadronic variables $z$ and $w$ introduced to describe the semileptonic
decay are redundant in this case, because the photon has $q^{2}=0;$ we have: 
\begin{eqnarray}
\qquad \qquad \qquad \qquad \qquad \qquad \qquad \qquad 1-z &=&\frac{1-x}{%
\left( 2-x\right) ^{2}}\approx 1-x,  \notag \\
w\, &=&\,\,2\,-\,x\,\,\approx \,\,1,
\end{eqnarray}
where $x\equiv x_{\gamma }.$ The variables $z$ and $x$ basically coincide in
the semi-inclusive region while the hadron energy is fixed, as anticipated
in the introduction. We therefore consider the $x$ spectrum. The improved
distribution we propose reads: 
\begin{equation}
\frac{d\Gamma }{dx}=\frac{G_{F}^{2}\alpha _{em}}{32\pi ^{4}}%
|V_{tb}V_{ts}^{\ast }|^{2}m_{b,pole}^{3}m_{b,\overline{MS}}^{2}\left(
m_{b}\right) \left[ \,Q\left( \mu _{b}\right) \,f\left( x\right) -\rho
^{\prime }\left( x\right) \right] +O\left( \alpha _{S}^{2}\right) ,
\label{trovata}
\end{equation}
where 
\begin{equation}
Q\left( \mu _{b}\right) \equiv |\widetilde{C}_{7}^{\left( 0\right) }\left(
\mu _{b}\right) |^{2}+\frac{\alpha _{S}\left( \mu _{b}\right) }{2\pi }\func{%
Re}\left\{ \widetilde{C}_{7}^{\left( 0\right) }\left( \mu _{b}\right) ^{\ast
}\left[ \widetilde{C}_{7}^{\left( 1\right) }\left( \mu _{b}\right)
+\sum_{i=1}^{8}\widetilde{C}_{i}^{\left( 0\right) }\left( \mu _{b}\right)
\left( r_{i}+\widetilde{\gamma }_{i7}^{\left( 0\right) }\log \frac{m_{b}}{%
\mu _{b}}\right) \right] \right\} .
\end{equation}
The ``remainder'' function $\rho ^{\prime }\left( x\right) $ is the
derivative of: 
\begin{equation}
\rho \left( x\right) \equiv \frac{\alpha _{S}\left( \mu _{b}\right) }{\pi }%
\sum_{i\leq j}^{1,8}\widetilde{C}_{i}^{\left( 0\right) }\left( \mu
_{b}\right) \,\widetilde{C}_{j}^{\left( 0\right) }\left( \mu _{b}\right)
\,f_{ij}\left( 1-x\right) .
\end{equation}
The latter vanishes at the endpoint: 
\begin{equation}
\rho \left( x\right) \rightarrow 0\qquad \mathrm{for}\qquad x\rightarrow 1,
\end{equation}
because the functions $f_{ij}\left( 1-x\right) $ all vanish in the endpoint: 
\begin{equation}
f_{ij}\left( 1-x\right) \rightarrow 0\qquad \mathrm{for}\qquad x\rightarrow
1.
\end{equation}
The function $\rho ^{\prime }\left( x\right) $ has at most a logarithmic
singularity $\ \,\sim \log \left( 1-x\right) $ for $x\rightarrow 1.$ The
quantities $\widetilde{C}_{i}^{\left( 0\right) }$ and $\widetilde{C}%
_{i}^{\left( 1\right) }$ are the leading and next-to-leading contributions
to the effective coefficient functions: 
\begin{equation}
\widetilde{C}_{i}\left( \mu \right) =\widetilde{C}_{i}^{\left( 0\right)
}\left( \mu \right) +\frac{\alpha _{S}\left( \mu \right) }{4\pi }\widetilde{C%
}_{i}^{\left( 1\right) }\left( \mu \right) +\cdots .
\end{equation}
$r_{i}$ are complex constants depending on the ratio $m_{c}/m_{b}.$ The
definition of all the symbols can be found in \cite{misiak}.

\psfig{bbllx=90pt, bblly=335pt, bburx=650pt, bbury=740pt,
file=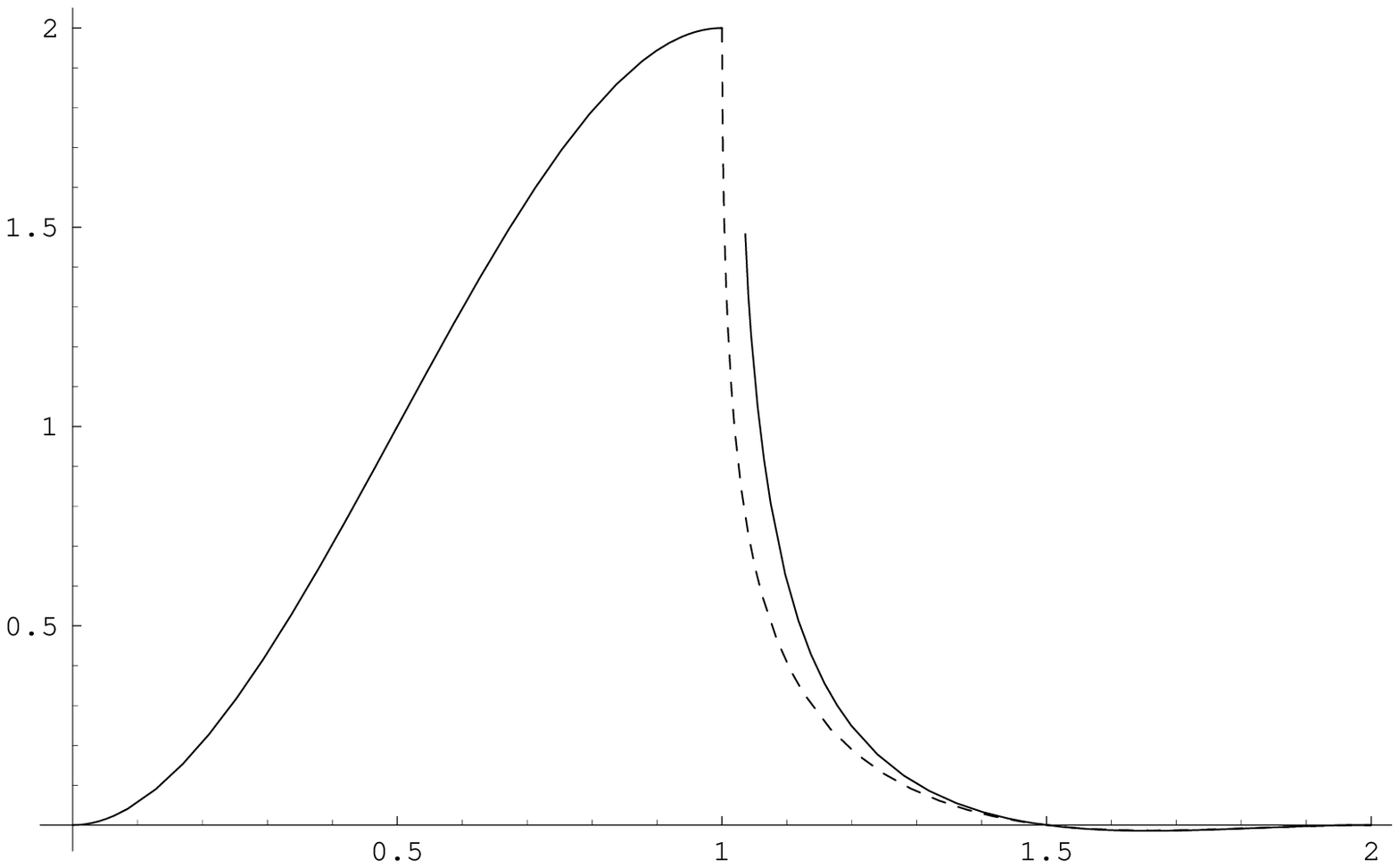, height=9cm, width=13cm} \vspace{-1.5cm}

\noindent \textit{Fig.~3: Plot of the energy spectrum in leading logarithmic
approximation. For }$w>1$\textit{, the dashed line corresponds to frozen
coupling, the continuous line to running coupling. In the latter case, the
distribution is not well defined close to the point }$w=1$\textit{\ because
of Landau-pole effects.}

\noindent To avoid large logarithms in the matrix elements, one has to take: 
\begin{equation}
\mu _{b}=O\left( m_{b}\right) .
\end{equation}
The long-distance effects are contained in the function $f\left( x\right) ,$%
which receives all the non-perturbative contributions (in leading twist)
related to Fermi motion. The main point is that $f\left( x\right) $ is the
same function $f\left( z\right) ,$ which was introduced to factorize the
long-distance effects in the semileptonic decay (cf. eq.\thinspace (\ref
{bellissima})). \ Therefore we explicitly verify the universality of the
long-distance contributions in (\ref{one}), once they are factorized by
means of a function of $z.$ The function $\rho ^{\prime }\left( x\right) $
is instead a short-distance contribution, specific of the process (\ref{rare}%
).

The proof of eq.\thinspace (\ref{trovata}) is trivial. The integrated
distribution in the upper part of the photon spectrum, 
\begin{equation}
\Gamma _{up}\left( x\right) \equiv \int_{x}^{1}\frac{d\Gamma }{dx^{\prime }}%
dx^{\prime },
\end{equation}
\ given in eq.\thinspace (30) of \cite{misiak}, can be written, up to terms
of order $\alpha _{S}^{2\text{ }},$ as: 
\begin{eqnarray}
\Gamma _{up}\left( x\right)  &=&\frac{G_{F}^{2}\alpha _{em}}{32\pi ^{4}}%
m_{b,pole}^{3}m_{b,\overline{MS}}^{2}\left( m_{b}\right) |V_{tb}V_{ts}^{\ast
}|^{2} \\
&&\left\{ Q\left( \mu _{b}\right) \left[ \theta \left( 1-x\right) -\frac{1}{2%
}A_{1}\alpha _{S}\left( \mu _{b}\right) \log ^{2}\left( 1-x\right)
+B_{1}\alpha _{S}\left( \mu _{b}\right) \log \left( 1-x\right) \right] +\rho
\left( x\right) \right\} +O\left( \alpha _{S}^{2}\right) .  \notag
\end{eqnarray}
One then takes a derivative with respect to $x.$ The requirement that the
spectrum is non-singular at $x=1$ --- or equivalently, that the total rate
is correctly reproduced\footnote{%
We do not consider here the problem of the infrared singularities for $%
x\rightarrow 0,$ related to soft-photon emission.} --- transforms the
distributions in plus distributions and one obtains eq.\thinspace (\ref
{trovata}).

\section{Conclusions}

We have performed factorization and threshold resummation of the most
general distribution in the semileptonic $b\rightarrow u$ decay. This has
been achieved with a proper choice of the kinematic variables, 
\begin{equation}
w\equiv \frac{2E_{X}}{m_{B}}\qquad \mathrm{and\qquad }z\equiv 1-\frac{%
m_{X}^{2}}{4E_{X}^{2}},
\end{equation}
which disentangle the two different logarithmic structures occurring in the
process: 
\begin{equation}
1):\,\,\log w\qquad \mathrm{and\qquad }2):\,\,\log \left( 1-z\right) .
\end{equation}
The first structure has been related to the infinite mass limit of the $b$
quark, 
\begin{equation}
m_{B}\rightarrow \infty ,
\end{equation}
while the second one has been related to the infinite energy limit of the
final hadronic system, 
\begin{equation}
E_{X}\rightarrow \infty .
\end{equation}
The long-distance effects --- both perturbative and non-perturbative ---
have been relegated to a universal factor $f\left( z\right) ,$ depending
only on $z.$ This function is related to the shape function $\varphi \left(
k_{+}\right) $ by a short-distance coefficient function. In essence, in our
approach, the hard scale of the process is the hadronic energy $E_{X}$ in
the $B$ rest frame, and not the $B$ meson mass $m_{B}.$

We have presented simple analytical expressions for a few distributions,
resummed \ to leading logarithmic accuracy and we have shown that the shape
function can be directly determined by measuring the distribution in $z$ or
in $z$ and $w.$

We have computed the resummed hadron energy spectrum, which exhibits a
``Sudakov shoulder'' in the point 
\begin{equation}
E_{X}=\frac{m_{B}}{2},
\end{equation}
i.e. in the middle of the allowed kinematical domain $\left( 0\leq E_{X}\leq
m_{B}\right) .$ The spectrum very close to this point is non-perturbative
and is proportional to the shape function. This implies that, at least in
principle, an accurate measure in this region may lead to another
independent determination of the shape function. It is a non-trivial fact
that the same long-distance effects appear at the boundary of the phase
space in the $z$-distribution while they appear inside the allowed
kinematical domain in the energy spectrum. The computation of other resummed
distributions in LO or NLO by integrating eq.\thinspace (\ref{bellissima})
is straightforward. A crucial point in our analysis is that there is a
single source of large logarithms in any semi-inclusive distribution in
heavy-flavour decay.

Finally, we have presented an improved formula for the photon spectrum in
the rare decay (\ref{rare}), which takes into account soft-gluon resummation
and the effects of the subleading operators. It has been explicitly shown
that the same function $f\left( z\right) $ factorizes the long-distance
effects in the semileptonic and in the rare decay.

We believe that our formalism sets a rather general and rigorous scheme for
the separation of perturbative from non-perturbative effects in
semi-inclusive heavy-flavour decays and allows for a simple analysis of the
experimental data of many different distributions.

\begin{center}
Acknowledgements
\end{center}

I wish to express particular thanks to S. Catani and M. Ciuchini for
suggestions. I also acknowledge discussions with S. Petrarca.


\begin{thebibliography}{99}
\bibitem{congiulia2}  U. Aglietti and G. Ricciardi, Nucl. Phys. B 587, 363
(2000).

\bibitem{dimenticato}  M. Shifman and A. Vainshtein, SJNP 45, 292 (1987); D.
Politzer and M. Wise, Phys. Lett.\thinspace B 206, 681 and B~208, 504 (1988).

\bibitem{ndf}  M. Neubert and F. De Fazio, JHEP 06, 017 (1999).

\bibitem{nostri}  G. Altarelli, N. Cabibbo, G. Corb\'{o}, L. Maiani and G.
Martinelli, Nucl. Phys. B 208, 365 (1982).

\bibitem{generale}  I. Bigi, M. Shifman, N. Uraltsev and A. Vainshtein,
Phys. Rev. Lett. 71, 496 (1993); Int. J. Mod. Phys. A 9, 2467 (1994); A.
Manohar and M. Wise, Phys. Rev. D 49, 1310 (1994); M. Neubert, Phys. Rev. D
49, 3392 and 4623 (1994); T. Mannel and M. Neubert, Phys. Rev. D 50, 2037
(1994).

\bibitem{penultimo}  U. Aglietti, preprint CERN-TH/2001-035, hep-ph/0102138.

\bibitem{ultimo}  U. Aglietti, preprint CERN-TH/2001-050, hep-ph/0103002.

\bibitem{kuhn}  M.\thinspace Jezabek and J.\thinspace Kuhn, Nucl. Phys. B
320, 20 (1989).

\bibitem{limetto}  A. Falk, M. Luke and M. Savage, Phys. Rev. D 53, 2491
(1996).

\bibitem{civuole}  R. Akhouri and I.~Rothstein, Phys. Rev. D 54, 2349
(1996); G.~Korchemsky and G. Sterman, Phys.\thinspace Lett.\thinspace B 340,
96 (1994);

\bibitem{sirlin}  R. Behrends, R. Filkelstein and A. Sirlin, Phys. Rev. 101,
866 (1956).

\bibitem{sudshoul}  S. Catani and B. Webber, JHEP 9710, 005 (1997); Phys.
Lett. B 427, 377 (1998).

\bibitem{sumrule}  U. Aglietti, preprint CERN-TH/2000-309, hep-ph/0010251.

\bibitem{aggiunta}  A. Falk, M. Neubert and M. Luke, Nucl. Phys. B 388, 363
(1992).

\bibitem{civuole2}  A. Leibovich and I.~Rothstein, Phys. Rev. D 61, 074006
(2000); A. Leibovich, I. Low and I.~Rothstein, Phys. Rev. D 61, 053006
(2000), hep-ph/0001028 and hep-ph/0005124. \ 

\bibitem{kodtren}  J. Kodaira and L. Trentadue, preprint SLAC-PUB-2934 and
Phys. Lett. B 112, 66 (1982); S.\thinspace Catani, E.\thinspace D'Emilio and
L. Trentadue, Phys. Lett. B 211, 335 (1988).

\bibitem{cattren}  S. Catani and L. Trentadue, Nucl. Phys. B 327, 323 (1989).

\bibitem{mangano}  S. Catani, M. Mangano, L. Trentadue and P. Nason, Nucl.
Phys. B 478, 273 (1996).

\bibitem{thrust}  S. Catani, L. Trentadue, G. Turnock and B. Webber, Nucl.
Phys. B 407, 3 (1993).

\bibitem{citare}  A. Ali and C. Greub, Z. Phys. C 49, 431 and Phys. Lett. B
259, 182 (1991); A. Kapustin and Z. Ligeti, Phys. Lett. B 355, 318 (1995);
N. Pott, Phys. Rev. D 54, 938 (1996); C. Greub, T. Hurth and D. Wyler, Phys.
Lett. B 380, 385 and Phys. Rev. D 54, 3350 (1996).

\bibitem{inostri}  M. Ciuchini, E. Franco, L. Reina, G. Martinelli and L.
Silvestrini, Phys. Lett. B 316, 127 (1993); M.\thinspace Ciuchini,
E.\thinspace Franco, L.\thinspace Reina and L.\thinspace Silvestrini, Nucl.
Phys. B 421, 41 (1994); M. Ciuchini, E. Franco, L. Reina, G.\thinspace
Martinelli and L.\thinspace Silvestrini, Nucl. Phys. B 415, 308 (1994).

\bibitem{misiak}  K. Chetyrkin, M. Misiak and M. Munz, Phys. Lett. B 400,
206 (1997); Erratum B 425, 414 (1998).
\end{thebibliography}
\end{document}